\documentclass[]{fairmeta}
\usepackage[utf8]{inputenc}

\usepackage{amsmath}
\usepackage{array}
\usepackage{enumitem}
\usepackage{float}
\usepackage{tabularx}
\usepackage{tikz}
\usetikzlibrary{arrows.meta,positioning,shapes.geometric,fit}

\setcitestyle{numbers,square,sort&compress}

\setlist[itemize]{leftmargin=*,itemsep=2pt,topsep=4pt}
\newcolumntype{Y}{>{\raggedright\arraybackslash}X}
\makeatletter
\newcommand\authorNewLine[2][]{%
  \addtolist[#1]{#2}{\authorlist}{\authorformat}{\\ }%
}

\makeatother

\labheader{AGI Lab | Westlake University}
\logo{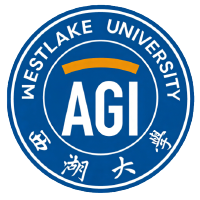}

\title{AppAgent-Claw: CLI Is All You Need for GUI Automation}
\author[1]{Zhixue Song}
\author[1]{Zhiheng Zhang}
\authorNewLine[1,2]{Yi Song}
\author[1*]{Chi Zhang}

\affiliation[1]{AGI Lab, Westlake University}
\affiliation[2]{Teeni AI}

\contribution[*]{Corresponding author}

\abstract{The OpenClaw platform provides a practical foundation for automation through its skill-oriented architecture, organizing external capabilities into lightweight, reusable components that can be invoked efficiently through a command-line interface (CLI). However, a significant bottleneck remains: many real-world tasks are confined to graphical user interfaces (GUIs) with no stable API available. While LLM-based GUI agents offer generality, their reliance on repeated live model inference makes them too slow, costly, and inconsistent to serve as efficient OpenClaw skills. In this paper, we present AppAgent-Claw, a demonstration-driven system that converts GUI workflows into reliable, reusable skills without runtime inference. By following a ``record-once, replay-many'' paradigm, the system captures rich contextual metadata to facilitate robust execution. It employs a layered localization strategy to handle visual shifts and a validation-coupled execution model to ensure intended on-screen effects. AppAgent-Claw provides a practical, efficient, and diagnosable solution for integrating GUI-bound tasks into the OpenClaw ecosystem.}

\metadata[Project Page]{\url{https://github.com/Westlake-AGI-Lab/AppAgent-Claw}}

\begin{document}
\maketitle

\section{Introduction}
OpenClaw has emerged as an especially compelling agent platform for practical automation~\cite{openclawgithub2026}. At its core, OpenClaw organizes external capabilities as lightweight, composable skills that can be bundled with the system. Once a capability is packaged as a skill, it can be invoked naturally by the agent  through a simple command-line interface  and reused seamlessly across many workflows without additional integration effort.  This skill-oriented design emphasizes fast invocation, low execution overhead, and reliable reuse, making it a practical foundation for deploying automation capabilities in real-world settings.

Yet despite these strengths, a critical gap remains. A large fraction of real-world software functionality is still exposed primarily through graphical user interfaces, with no stable API or command-line interface for the operations users perform every day. This means that even when OpenClaw reasons correctly and plans effectively, it may still be unable to act because the software it needs to operate is only reachable through a GUI that the agent has no reliable way to control. For a platform whose value lies in automating real-world tasks, this represents a fundamental ceiling on what the agent can actually accomplish. Closing this gap is the central motivation of this work.

The most prominent recent direction for GUI automation is the GUI agent, such as AppAgent~\cite{zhang2025appagent,jiang2025appagentxevolvingguiagents}: systems powered by large language models that perceive live screenshots and reason about actions at every step. While these agents demonstrate impressive generality, their open-world approach is fundamentally expensive and slow, requiring repeated model inference at every invocation and offering no guarantee of consistent results across runs. This stands in direct tension with what an OpenClaw skill is expected to be: fast, consistent, and lightweight enough to be called on demand. A GUI agent brings the full cost of open-world perception to every invocation, making it a poor fit for the skill model that OpenClaw depends on.

We present AppAgent-Claw, a GUI automation skill for OpenClaw motivated by a simple but practical goal: to enable OpenClaw to learn a GUI workflow through a demonstration, so that the agent can invoke it efficiently and reliably as a reusable skill, without repeated model inference at execution time.
The system follows a record-once, annotate-once, replay-many pattern across three phases. In the recording phase, AppAgent-Claw captures not only the user's actions, but also the rich visual and contextual information surrounding each interaction. This preserved context is what enables robust replay later, without requiring any live model inference at execution time. In the annotation phase, the goal is to transform a raw recording into a form that an agent can understand and reuse. The demonstration is enriched with semantic descriptions and parameterized inputs, such as variable text fields that differ across runs, so that the resulting workflow is not a fixed replay of one specific instance, but a general, named skill that can be invoked flexibly across different contexts. In the replay phase, the central challenge is finding the right target on screen when visual conditions have shifted slightly since recording. Rather than relying on a single brittle strategy, AppAgent-Claw attempts localization in layers, starting from a precise local match and widening the search progressively if earlier attempts fail. Execution is further coupled with post-action validation that confirms the intended effect has actually occurred, treating a confirmed outcome as the true measure of success rather than a dispatched action.

The result is a system whose reliability comes not from any single powerful model, but from preserved context, layered fallback, and explicit observability: an approach that degrades gracefully, makes failures diagnosable, and fits naturally into OpenClaw's lightweight skill model.
In this paper, we propose AppAgent-Claw, a practical skill for converting a user demonstration into a reliable, reusable GUI workflow on the OpenClaw platform. The contributions of this work are as follows. A demonstration-driven recording protocol captures rich visual and contextual information during a user walkthrough, providing the grounding necessary for robust replay without live model inference. A layered localization strategy searches for targets progressively from local to global, degrading gracefully when visual conditions shift between recording and replay. A validation-coupled execution model treats a confirmed on-screen effect, rather than a dispatched action, as the true measure of success, making failures recoverable and diagnosable.
\section{Related Work}
\textbf{Rule Based Automation, GUI Agents, and Positioning.} Classical GUI automation and RPA emphasize deterministic execution through coordinates, UI trees, trigger rules, and hand authored workflows, offering low runtime cost and strong repeatability but often relying on brittle selectors or fixed state assumptions~\cite{van2018robotic,syed2020rpa}. Systems such as Sikuli show that screenshot-based visual anchors can provide a practical middle ground between rigid coordinate replay and richer interface understanding~\cite{yeh2009sikuli}. Recent GUI agents broaden the goal to multimodal, instruction-guided interaction across interfaces. On mobile devices, AppAgent, AppAgentX, Mobile Agent, and AutoDroid study visual grounding and autonomous operation~\cite{zhang2025appagent,jiang2025appagentxevolvingguiagents,wang2024mobileagent,wen2024autodroid}; on the web, WebVoyager, AutoWebGLM, and OpenWebVoyager explore realistic navigation with multimodal perception and iterative improvement~\cite{he2024webvoyager,lai2024autowebglm,he2025openwebvoyager}; and on desktop or general computer control settings, ASSISTGUI, UFO, OS Copilot, ScreenAgent, Agent S, Agent S2, Cradle, and Scaling Agents for Computer Use investigate planning, reflection, grounding, and long horizon execution across applications~\cite{gao2023assistgui,zhang2024ufo,wu2024oscopilot,niu2024screenagent,Agent-S,Agent-S2,tan2024cradle,Agent-S3}. Compared with these systems, AppAgent-Claw is narrower in scope: it targets repeated foreground workflows and turns them into reusable automation skills rather than pursuing broad open domain autonomy.

\textbf{Grounding, Instruction Following, and Computer Control.} Reliable GUI execution depends on mapping language and screenshots to concrete UI targets. SeeClick and CogAgent strengthen screenshot understanding and element grounding, while Rico provides a widely used source of UI structure~\cite{cheng2024seeclick,hong2024cogagent,deka2017rico}. Earlier work on following instructions and computer control, like mapping natural language instructions to mobile UI action sequences, learning control policies from human computer interaction data, and zero shot reflective control on browser tasks, likewise highlights the difficulty of turning high level intent into dependable low level actions~\cite{li2020mapping,humphreys2022computercontrol,li2023structuredreflection}. This directly motivates AppAgent-Claw's design choice to preserve anchor crops, context images, monitor geometry, and window metadata during recording rather than relying on single shot open world perception at replay time.

\textbf{Demonstration Driven Automation and Workflow Reuse.} Beyond autonomous GUI agents, a long line of work studies how systems can learn procedures from user demonstrations and turn them into reusable workflows. Programming by demonstration and interface agent research, from Metamouse and EAGER to scripting graphical applications by demonstration, Watch What I Do, and Your Wish Is My Command, emphasizes capturing human actions, abstracting them into higher level procedures, and reusing them in future tasks~\cite{maulsby1989metamouse,cypher1991eager,myers1998scripting,cypher1993watch,lieberman2001wish}. More recent end-user systems such as SUGILITE show that demonstration remains a practical route to interface automation, especially when combined with richer multimodal interaction~\cite{li2017sugilite}. Work on visual generalization likewise studies how reusable procedures can transfer beyond the exact original instance~\cite{amant2000visual}. At the learning level, imitation learning perspectives treat demonstration as a source of structured supervision for action sequencing and policy learning~\cite{hussein2017imitation}. AppAgent-Claw is closest to this direction in spirit, but differs in its systems focus: rather than learning an open ended control policy, it preserves replay critical visual and window context from a concrete foreground workflow and reuses that context through constrained matching, fallback, and validation. This makes the system better understood as a practical workflow reuse skill than as a general purpose autonomous GUI agent.

\section{Method}

In this section, we present AppAgent-Claw, a demonstration-driven GUI automation skill designed to address the lack of lightweight, reliable GUI control in the OpenClaw platform. We begin with an overview of the system architecture in Section~\ref{sec:system-overview}, followed by the end-to-end skill workflow in Section~\ref{sec:skill workflow}. We then describe the core stages of the system, including recording (Section~\ref{sec:recording stage}), annotation (Section~\ref{sec:annotation}), layered replay localization (Section~\ref{sec:replay}), and validation-coupled execution (Section~\ref{sec:execution}).

\subsection{System Overview}
\label{sec:system-overview}
At a high level, the system consists of recording, annotation, replay localization, action execution, and result persistence. Table~\ref{tab:modules} summarizes the main functional components.

\begin{table}[htbp]
\centering
\small
\begin{tabularx}{\linewidth}{@{}lY@{}}
\toprule
\textbf{Module} & \textbf{Role} \\
\midrule
Recording controller & Controls interactive recording sessions and visible recording status. \\
Event aggregator & Aggregates keyboard and mouse events into semantic steps. \\
Screen capture and context extraction & Captures target crops, surrounding context, search regions, and monitor geometry metadata. \\
Flow annotation & Adds flow descriptions, step semantics, and candidate text inputs to the flow definition. \\
Replay coordinator & Prepares window context, resolves targets, executes actions, validates results, and records outcomes. \\
Target resolver & Performs template matching and relative coordinate fallback during replay. \\
Action executor & Implements clicking, scrolling, typing, hotkeys, waiting, and window restoration. \\
Storage layer & Stores recording assets, execution artifacts, and structured flow data. \\
\bottomrule
\end{tabularx}
\caption{Main functional components of AppAgent-Claw and their responsibilities across the workflow. The table shows how the system moves from interactive recording and annotation to replay execution and artifact storage.}
\label{tab:modules}
\end{table}

Runtime artifacts are separated into reusable recording assets and execution specific diagnostics. The flow definition stores relative references rather than host specific absolute paths. This keeps artifact lookup self contained within one skill directory and cleanly separates reusable assets from later execution traces.

\begin{figure}[H]
\centering
\small
\includegraphics[width=\linewidth]{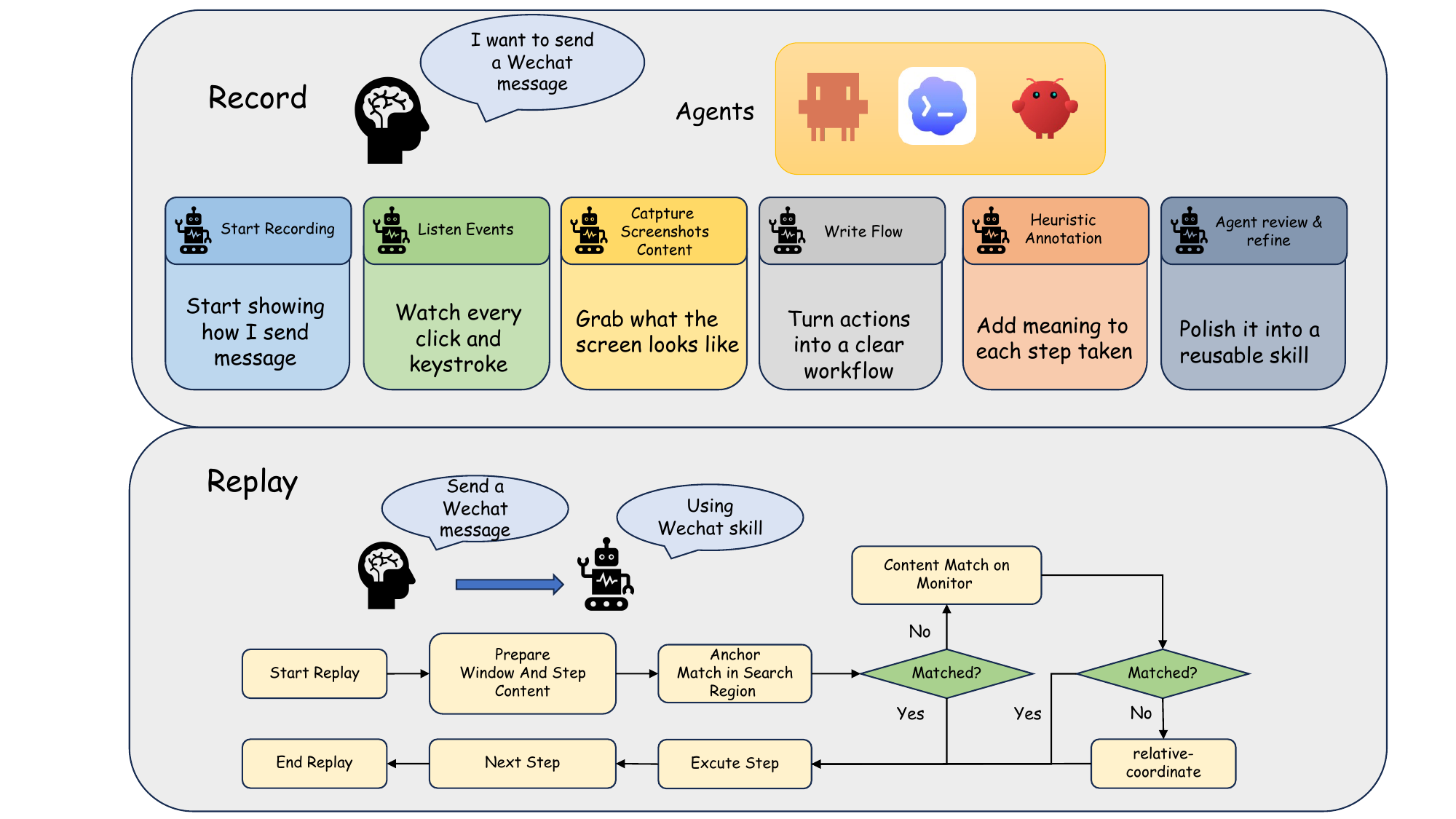}

\caption{
Overview of the AppAgent-Claw workflow. The top panel shows recording and annotation, where user interactions are captured, organized into workflows, and refined into reusable skills. The bottom panel shows replay, where the system executes the workflow using layered target localization through anchor matching, context matching, and relative coordinate fallback.
}
\label{fig:workflow}
\end{figure}

\subsection{Skill Workflow}
\label{sec:skill workflow}
The skill follows the pattern \textit{record once, annotate once, replay many times}. Figure~\ref{fig:workflow} summarizes how a raw foreground demonstration is transformed into a reusable skill and later executed.

Concretely, the recording half of the workflow proceeds through six stages. A user first demonstrates the target task in the foreground application. The system then listens to mouse and keyboard events, captures local screenshots and surrounding context for click related actions, and writes the observed sequence into a structured flow definition. Automatic annotation adds draft descriptions and candidate text inputs, but this output is only an initial semantic pass. Before the flow is treated as reusable, a downstream reviewer, either human or agent, inspects the recorded steps and screenshots, verifies the intended task, refines the descriptions, and confirms which text inputs can be changed safely in future runs. The recorded action sequence itself remains fixed during this review.

The replay half begins when an agent invokes the reviewed skill, optionally with runtime text inputs. For each step, the system prepares the relevant window context and resolves the target before execution. As shown in Figure~\ref{fig:workflow}, target resolution follows a layered policy. The system first attempts a local anchor match within the recorded search region. If that is insufficient, it expands to a broader context match over the recorded monitor area. Only when both visual cues fail does it fall back to monitor relative coordinates. After a target is resolved, the action is executed, the resulting on screen effect is validated, and the workflow advances to the next step. In this way, replay depends on both preserved visual context from recording and the review pass that turns a raw demonstration into a semantically reusable skill.

\subsection{Data Protocol}
\label{sec:data protocol}
Before detailing the three core stages, we describe the data protocol that underlies all of them, as it defines the information contract between recording, annotation, and replay.

The main data object is a structured flow definition that stores top level fields such as \texttt{schema\_version}, \texttt{name}, \texttt{platform}, \texttt{created\_at}, \texttt{app\_context}, \texttt{inputs}, \texttt{annotation}, and \texttt{steps}. Among them, \texttt{steps} is the primary execution carrier during replay. Click related steps are more constrained than other actions because the system must find targets again later instead of replaying fixed pixel positions.

\begin{table}[H]
\centering
\small
\begin{tabularx}{\linewidth}{@{}lY@{}}
\toprule
\textbf{Field} & \textbf{Meaning} \\
\midrule
\texttt{monitor} & Absolute geometry of the monitor used during recording. \\
\texttt{target} & Absolute target coordinates and relative monitor coordinates. \\
\texttt{locator} & Anchor crop, context crop, local search region, and matching threshold used during replay. \\
\texttt{retry} & Maximum attempts and whether relative coordinate fallback is allowed. \\
\texttt{validation} & Post action validation mode and timeout settings. \\
\texttt{window\_context} & App name, title, position, and size of the owning window. \\
\bottomrule
\end{tabularx}
\caption{Main metadata stored for click related steps in the flow definition. Together, these fields provide the context needed for localization, retry control, validation, and window restoration during replay.}
\label{tab:click-fields}
\end{table}

Absolute pixel coordinates are tied to a specific monitor geometry and can become invalid when the display configuration changes. To improve robustness across different environments, we normalize each target location into monitor-relative coordinates:
\[
\texttt{rel\_x} = \frac{x - \texttt{monitor.left}}{\texttt{monitor.width}},
\qquad
\texttt{rel\_y} = \frac{y - \texttt{monitor.top}}{\texttt{monitor.height}}.
\]
When template matching fails, replay can project the target into the current monitor geometry as a constrained fallback, rather than reusing historical absolute pixels.

The storage design separates reusable assets from diagnostics. Assets needed for future replay remain attached to the recorded flow, while each execution produces an isolated set of logs and debug traces. This split avoids cross contamination between reusable artifacts and troubleshooting evidence.

\subsection{Action Space}
\label{sec:action space}
We also enumerate the supported action space, which determines what user interactions can be captured and replayed by the system. The supported action types are listed in Table~\ref{tab:actions}.

\begin{table}[htbp]
\centering
\small
\begin{tabular*}{0.72\linewidth}{@{\extracolsep{\fill}}ll@{}}
\toprule
\textbf{Action} & \textbf{Description} \\
\midrule
\texttt{move} & Move cursor to a target coordinate. \\
\texttt{click} & Left single click. \\
\texttt{double\_click} & Left double click. \\
\texttt{right\_click} & Right single click. \\
\texttt{long\_press} & Left button long press. \\
\texttt{right\_long\_press} & Right button long press. \\
\texttt{scroll} & Vertical or horizontal scrolling. \\
\texttt{type\_text} & Text input. \\
\texttt{hotkey} & Single key or key combination. \\
\texttt{wait} & Explicit delay. \\
\bottomrule
\end{tabular*}
\caption{Action types defined in the current workflow schema and supported during replay. The table covers pointer actions, text input, hotkeys, scrolling, and explicit waits.}
\label{tab:actions}
\end{table}

Schema support is broader than stable recording output. The current recording path reliably emits click, double click, right click, long press, scroll, text input, hotkey, and wait. Although \texttt{move} exists in the schema and execution layer, it is not a primary output of automatic recording.

\subsection{Recording Stage: From Event Stream to Structured Steps}
\label{sec:recording stage}
The recording stage converts raw keyboard and mouse activity into structured steps. Rather than emitting every low level event directly, it performs semantic aggregation before producing the final action sequence.

This aggregation depends on short term state that includes a continuous text buffer, pending mouse down status, active modifier set, last click metadata, and the latest stable window context. Together, these signals determine whether a raw event becomes a single click, double click, long press, committed text input, or explicit wait.

\begin{table}[htbp]
\centering
\small
\begin{tabularx}{\linewidth}{@{}l>{\raggedleft\arraybackslash}p{2.2cm}Y@{}}
\toprule
\textbf{Parameter} & \textbf{Default} & \textbf{Purpose} \\
\midrule
\texttt{wait\_threshold\_seconds} & 0.8s & Insert \texttt{wait} when the event interval exceeds the threshold. \\
\texttt{double\_click\_interval\_seconds} & 0.35s & Temporal window for double click detection. \\
\texttt{double\_click\_distance\_px} & 4px & Spatial tolerance for double click detection. \\
\texttt{long\_press\_threshold\_seconds} & 0.5s & Hold duration threshold for long press. \\
\texttt{long\_press\_distance\_px} & 8px & Motion tolerance during long press. \\
\texttt{ime\_commit\_delay\_seconds} & 0.05s & Short delay after IME commit. \\
\texttt{ime\_commit\_poll\_timeout\_seconds} & 0.35s & Max polling time for focused text change. \\
\bottomrule
\end{tabularx}
\caption{Thresholds used to aggregate raw keyboard and mouse events into structured recording steps. They control wait insertion, click disambiguation, long press detection, and IME commit handling.}
\label{tab:thresholds}
\end{table}

Click related sampling is central to replay quality. The system does not crop templates immediately at mouse down. Instead, it caches a full monitor screenshot first, then crops \texttt{anchor} and \texttt{context} after mouse up and action classification. This improves click type disambiguation and reduces the chance of preserving transient intermediate UI states as anchors.

\begin{table}[htbp]
\centering
\small
\begin{tabular*}{0.50\linewidth}{@{\extracolsep{\fill}}lc@{}}
\toprule
\textbf{Sampling Item} & \textbf{Default} \\
\midrule
\texttt{anchor\_width} & 96 \\
\texttt{anchor\_height} & 96 \\
\texttt{context\_width} & 240 \\
\texttt{context\_height} & 160 \\
\texttt{search\_padding} & 96 \\
\bottomrule
\end{tabular*}
\caption{Default crop sizes and search padding in pixels used when recording click related targets. These values define the anchor crop, the broader context crop, and the local search region preserved for later replay.}
\label{tab:sampling}
\end{table}

In multi monitor settings, the system first resolves which monitor contains the click and stores both absolute screen coordinates and monitor relative coordinates. This does not guarantee generalization across layouts, but it is more consistent with the target assumption of \textit{same machine, similar layout} than hard coded absolute pixels.

When available, text recording also captures semantic result text rather than only key sequences. The system attempts to read \texttt{AXFocusedUIElement}~\cite{apple2023accessibility} and \texttt{AXValue}~\cite{apple_axvalue}, then observes focused text changes after IME commit. This helps separate final inserted text from intermediate key events, especially for CJK input and candidate selection. If the target application does not expose reliable accessibility text values, such as custom drawn controls, behavior degrades to coarser key sequence capture.

Window context is also collected during recording, using a fallback order of focused window, main window, and then first window. This improves the robustness of later window restoration.

\subsection{Annotation and Parameterized Text After Recording}
\label{sec:annotation}
After recording ends, the system runs automatic annotation. The annotator generates initial metadata for the structured flow definition, including flow level descriptions, step descriptions, candidate \texttt{inputs}, and \texttt{text\_policy} for relevant \texttt{type\_text} steps. This stage produces a draft semantic layer rather than a final interpretation.

Recording is therefore not considered complete when the CLI exits. A downstream reviewer, either human or agent, examines the saved flow definition, the screenshots for each step, step order, and optional images of the full flow. The reviewer confirms what the flow actually accomplishes, why each step exists, and which text content can be parameterized safely.

During this review, the recorded action sequence should not be edited, reordered, inserted, or removed. If action capture is wrong, the preferred remedy is to record again rather than patch metadata around the error. The goal of review is semantic clarity, so that another agent can understand intent from metadata and images alone.

Parameterized text follows a two stage policy: automatic candidate generation plus reviewer confirmation. Candidate detection considers whether submit hotkeys follow, whether clicks follow, whether the text appears free form, whether waits appear after submission, and whether this is the final \texttt{type\_text}. It also considers whether the application context resembles messaging or commenting scenarios. Final parameterization still requires full visual context and downstream anchor stability checks.

Only text that behaves like final message content should typically remain parameterized. Text that affects later locator stability, such as search keys or conditions that define state, should remain fixed. Therefore, the input specification is not raw recording output; it is a semantic layer produced jointly by heuristic annotation and review.

\subsection{Replay Localization: Template First, Relative Coordinates Last}
\label{sec:replay}
Replay's hardest problem is target resolution, not low level action emission. For click related steps, replay first prepares the recorded window context and then invokes the resolver to locate the target.

\begin{table}[htbp]
\centering
\small
\begin{tabularx}{\linewidth}{@{}lYl@{}}
\toprule
\textbf{Layer} & \textbf{Search Target and Scope} & \textbf{Purpose} \\
\midrule
Layer 1 & Match the anchor crop within the recorded local search region. & Fast and precise local localization. \\
Layer 2 & Match the broader context crop on the full recorded monitor area. & Recover when local anchor is unstable. \\
Layer 3 & Project monitor relative coordinates to current monitor geometry. & Constrained fallback when template matching fails. \\
\bottomrule
\end{tabularx}
\caption{Three stage localization policy used to resolve click related targets during replay. The resolver first attempts local anchor matching, then broader context matching, and finally monitor relative fallback.}
\label{tab:resolver-layers}
\end{table}

The default template threshold is 0.92. Matching is based on OpenCV~\cite{bradski2000opencv} \texttt{matchTemplate}, but method selection is data dependent: when either template or search image has zero grayscale standard deviation, the resolver switches to \texttt{TM\_SQDIFF\_NORMED}~\cite{brunelli2009template}; otherwise it uses \texttt{TM\_CCOEFF\_NORMED}. This avoids low discriminability failure modes on nearly solid color patches.

The resolver also performs quality filtering before matching. Templates with grayscale standard deviation $\leq 4.0$ and maximum RGB channel dynamic range $\leq 16$ are considered limited information templates (e.g., blank or low texture areas). Such strategies are skipped directly, and skip reasons are written into resolve attempts.

Only if both anchor and context fail, and \texttt{retry.fallback\_to\_relative} is true, does the system use relative coordinate fallback. Even this fallback is restricted to the recorded monitor and not an unconditional click on historical absolute pixels.

In debug mode or on terminal failure, resolver artifacts such as search crops and match visualizations are persisted as execution diagnostics. On normal successful paths, these heavy images remain in memory to avoid unnecessary disk noise.

\subsection{Execution, Validation, and Observability}
\label{sec:execution}
Once a target is resolved, the executor performs primitive actions. This layer intentionally avoids additional reasoning and returns unified \texttt{ActionResult} objects with success state, elapsed time, error code, and detailed metadata. The goal is to keep replay behavior observable and to support structured diagnosis when a run fails.

\begin{table}[htbp]
\centering
\small
\begin{tabularx}{\linewidth}{@{}lY@{}}
\toprule
\textbf{Capability} & \textbf{Mechanism} \\
\midrule
Click and long press & Synthesizes single, double, right click, and hold actions. \\
Scrolling & Supports vertical and horizontal scrolling, optionally after cursor repositioning. \\
Hotkeys & Executes single keys and key combinations. \\
Waiting & Inserts explicit delays when required by the workflow. \\
Text input & Uses clipboard based paste to insert final content in one step. \\
Window preparation & Activates the target application and attempts best effort restoration of window geometry. \\
\bottomrule
\end{tabularx}
\caption{Primitive executor operations used during replay. In addition to user facing actions, the executor also handles window preparation before a step is executed.}
\label{tab:executor}
\end{table}

Clipboard based text input is chosen for two practical reasons. It is less sensitive to IME state, candidate interaction, and keyboard layout across applications. It also better reflects single shot insertion of final content. To reduce side effects, the executor attempts clipboard backup and restore, and reports restore outcome in execution details.

Replay robustness is reinforced by multiple safeguards beyond action emission:
\begin{itemize}
  \item \textbf{Window preparation}: if \texttt{window\_context} is present, the system attempts app/title/geometry restoration, but skips it when recorded frame assumptions are clearly inconsistent with target points.
  \item \textbf{Retry policy}: click related steps expose \texttt{max\_attempts} and \texttt{fallback\_to\_relative}, so one failure does not necessarily terminate the flow.
  \item \textbf{Posterior validation}: action dispatch and state effect are validated separately.
\end{itemize}

Supported validation modes are listed in Table~\ref{tab:validation}.

\begin{table}[htbp]
\centering
\small
\begin{tabularx}{0.84\linewidth}{@{}lY@{}}
\toprule
\textbf{Mode} & \textbf{Meaning} \\
\midrule
\texttt{none} & No additional validation. \\
\texttt{anchor\_present} & Confirm that the expected next step anchor appears after execution. \\
\texttt{anchor\_absent} & Confirm that the current anchor disappears after execution. \\
\bottomrule
\end{tabularx}
\caption{Post action validation modes used during replay. The table shows how the system checks whether execution produced the expected screen state change.}
\label{tab:validation}
\end{table}

The \texttt{anchor\_present} mode reflects a workflow oriented design: instead of checking again whether the just clicked widget still exists, it verifies whether the \emph{next step's} target has appeared. For anchor checks, polling defaults to 0.2s intervals within a default 2.0s window.

For \texttt{type\_text}, the system prefers focused text validation: read before input text, execute input, then poll every 0.2s (default 2.0s window) for expected content in the focused control. If accessibility text APIs are unavailable, validation degrades to a conservative fixed 1.0s wait rather than immediate failure.

The runtime also exposes a lightweight status window during recording and replay to indicate the current step index, action type, and step description. To avoid contaminating replay images, the window is hidden during screenshot capture and template matching, then restored immediately afterward. It is also configured not to intercept user interactions.

\subsection{Layered Safeguards}
Viewed as a whole, AppAgent-Claw's reliability does not stem from a single component, but from a set of layered safeguards operating across stages.

During recording, the system preserves local visual and window context, providing stable grounding for replay. This grounding enables layered localization, where replay progresses from precise anchor matching to broader context matching and constrained coordinate fallback, allowing minor visual variations to be absorbed rather than causing immediate failure.

During execution, posterior validation decouples action dispatch from outcome confirmation, ensuring that the system advances only when the intended effect is observed. At the same time, structured diagnostics make failures observable, explainable, and retryable.

Overall, the system degrades gracefully under perturbations, fails explicitly when assumptions are violated, and recovers predictably through structured retries. Reliability therefore arises from constraining and managing failure, rather than eliminating it.

\section{Experiments}
\label{sec:experiments}

\subsection{Experimental Setup}
We evaluate AppAgent-Claw in the same machine setting that the system is designed for. All recordings and replays are collected on a single macOS machine. We study two questions: (1) whether the system can replay real workflows reliably under baseline conditions, and (2) whether the layered resolver continues to work when visual conditions drift away from recording.

The benchmark contains five workflows spanning note taking, browsing, media, and messaging: creating a new note in Notes, opening a search tab in Safari, playing daily recommendations in NetEase Music, playing favorite songs in NetEase Music, and opening a prepared chat flow in WeChat. We report end-to-end success rate, per-step success rate, average latency, and layer hit rate over click-like steps only. For layer attribution, the denominator is the number of committed resolver calls, not the total number of steps.

\subsection{Replay Reliability and Layer Attribution}
We first replay each of the five workflows ten times under baseline conditions, yielding 50 runs in total. Table~\ref{tab:exp-baseline} summarizes the results.

\begin{table}[htbp]
\centering
\small
\setlength{\tabcolsep}{4pt}
\begin{tabularx}{\linewidth}{@{}Yrrrrrrrr@{}}
\toprule
\textbf{Workflow} & \textbf{Steps} & \textbf{Clicks} & \textbf{E2E} & \textbf{Step} & \textbf{Lat.} & \textbf{L1} & \textbf{L2} & \textbf{L3} \\
& \textbf{/run} & \textbf{/run} & \textbf{Succ.} & \textbf{Succ.} & \textbf{(s)} & \textbf{Hit} & \textbf{Hit} & \textbf{Hit} \\
\midrule
Notes: create new note & 10 & 4 & 10/10 & 100.0\% & 19.7 & 75.0\% & 0.0\% & 25.0\% \\
Safari: open search tab & 7 & 2 & 10/10 & 100.0\% & 11.3 & 100.0\% & 0.0\% & 0.0\% \\
NetEase Music: play daily recommendation & 11 & 6 & 10/10 & 100.0\% & 28.7 & 83.3\% & 16.7\% & 0.0\% \\
NetEase Music: play favorite songs & 11 & 6 & 10/10 & 100.0\% & 24.7 & 100.0\% & 0.0\% & 0.0\% \\
WeChat: send message to File Transfer Assistant & 13 & 3 & 10/10 & 100.0\% & 18.1 & 63.3\% & 16.7\% & 20.0\% \\
\midrule
\textbf{Overall} & \textbf{--} & \textbf{--} & \textbf{50/50} & \textbf{100.0\%} & \textbf{20.5} & \textbf{85.2\%} & \textbf{7.1\%} & \textbf{7.6\%} \\
\bottomrule
\end{tabularx}
\caption{Baseline replay results under the intended same machine setting. Steps/run and Clicks/run denote per workflow counts. E2E Succ. reports completed runs, Step Succ. reports per step success rate, Lat. reports mean latency in seconds, and L1 to L3 Hit report resolver hit rates over the 210 click related resolver calls.}
\label{tab:exp-baseline}
\end{table}

Under the intended same machine replay setting, the baseline result is highly reliable: all 50 runs complete successfully, every executed step succeeds, and no error code is emitted anywhere in the pipeline. The layer distribution further suggests that the resolver design is practically useful rather than redundant. Across all click like steps, Layer 2 accounts for 7.1\% of committed resolves and Layer 3 accounts for 7.6\%, so 14.7\% of successful localizations come from paths beyond the primary anchor match. This pattern is also workflow dependent. Static interfaces such as Safari and NetEase favorites remain almost entirely in Layer 1, while more dynamic surfaces such as WeChat and newly created Notes entries rely more often on Layers 2 and 3.

\subsection{Robustness under Visual and Content Perturbation}
We next evaluate whether the layered resolver provides graceful degradation once the replay condition is perturbed. We select three representative workflows from Notes, NetEase Music, and WeChat, and replay each of them under four conditions: baseline, system-wide dark mode, in-app zoom, and a different initial UI state. Each workflow-condition pair is run three times, for 36 runs in total.

\begin{table}[htbp]
\centering
\small
\setlength{\tabcolsep}{4pt}
\begin{tabularx}{\linewidth}{@{}Yrrrrrrr@{}}
\toprule
\textbf{Condition} & \textbf{Runs} & \textbf{E2E} & \textbf{Step} & \textbf{Lat.} & \textbf{L1} & \textbf{L2} & \textbf{L3} \\
& & \textbf{Succ.} & \textbf{Succ.} & \textbf{(s)} & \textbf{Hit} & \textbf{Hit} & \textbf{Hit} \\
\midrule
Baseline & 9 & 9/9 & 100.0\% & 22.3 & 74.4\% & 12.8\% & 12.8\% \\
Dark mode & 9 & 9/9 & 100.0\% & 22.3 & 0.0\% & 5.1\% & 94.9\% \\
Zoom & 9 & 9/9 & 100.0\% & 22.5 & 76.9\% & 15.4\% & 7.7\% \\
Different UI state & 9 & 9/9 & 100.0\% & 22.5 & 74.4\% & 15.4\% & 10.3\% \\
\bottomrule
\end{tabularx}
\caption{Replay results under four perturbed conditions, aggregated over three workflows with three runs in each condition. E2E Succ. reports completed runs, Step Succ. reports per step success rate, Lat. reports mean latency in seconds, and L1 to L3 Hit report resolver hit rates over the 39 click related resolver calls in each condition.}
\label{tab:exp-perturb}
\end{table}

Dark mode is the clearest stress case. The Layer 1 hit rate drops from 74.4\% to 0.0\%, indicating that light mode anchor templates no longer transfer once the interface appearance changes substantially. End to end success nevertheless remains 100.0\%, because Layer 3 absorbs most resolver calls and Layer 2 covers the remainder. This suggests that the fallback design is essential for maintaining replay once the primary visual cue becomes unreliable. By contrast, zoom and different initial state perturbations are milder. Layer 1 still resolves most targets in both settings, while Layers 2 and 3 absorb the remaining mismatches. Average latency stays nearly constant across all four conditions, from 22.3 to 22.5\,s, suggesting that fallback does not materially change overall workflow completion time in this benchmark.

Taken together, the two experiments cover 86 successful replays and 366 click like resolver calls, again with no emitted error code. These results support two observations within the intended operating setting. First, the system is reliable for repeated replay on the same machine. Second, its layered localization design yields graceful degradation rather than immediate failure when visual conditions drift moderately from the recording environment. We note that the reported layer hit rates are attribution statistics over successful resolver calls rather than removal based ablations. They show how often each layer is ultimately needed in this benchmark, but they do not isolate component contribution under removal.

\section{Discussion}

The central design choice of the system is not a particular matching algorithm, but a deliberate restriction of scope. Whereas recent GUI agents aim for broad task coverage in open environments, the present system targets repeated workflows that the user has already demonstrated once. This narrower setting sacrifices breadth, but gains controllability, inspectability, and repeatable execution. The experiments should therefore be interpreted as evidence of practical reliability in a constrained replay regime, not as a claim of broad open domain GUI autonomy.

This framing also helps explain why the system works in practice. Its robustness comes less from any single perception component than from preserving replay critical context and coupling execution with explicit recovery and validation. In this design, success is defined not only by sending an action, but by observing the intended on screen effect under structured checks. Packaging the workflow as a lightweight skill further supports reuse, because it turns a demonstrated interaction into a callable capability without requiring a full computer use stack for each invocation.

That said, the approach has clear operating boundaries. It assumes repeated use on the same machine, foreground interaction, and sufficiently similar monitor and window layouts for recorded context to remain informative. The current results therefore should not be read as evidence of robustness to arbitrary interface redesign, transfer across devices, or autonomous exploration in previously unseen applications. For this reason, the system is best viewed not as a replacement for general GUI agents, but as a practical workflow substrate that could later be combined with them. In that longer term direction, user demonstration would serve not only as replay input but also as supervision, enabling agents to learn reusable workflow structure and execute it with increasing autonomy.

\section{Conclusion}
This report presented AppAgent-Claw, a practical skill for converting a user demonstration into a reusable GUI workflow through recording, annotation, layered target localization, validation, and structured diagnostics. Rather than pursuing open domain desktop intelligence, the system focuses on reliable workflow reuse under constrained assumptions and shows that strong practical behavior can emerge from preserved context, controlled fallback, and explicit observability. More broadly, the project suggests that demonstration driven automation can serve not only as a useful endpoint in its own right, but also as a foundation for future agents that learn from demonstration and act with increasing autonomy.

\bibliographystyle{unsrtnat}
\bibliography{references}

\end{document}